\documentclass[preprint,prd,showpacs]{revtex4}
\usepackage{mathrsfs}
\usepackage{amsmath}
\usepackage{amsfonts}
\usepackage{amssymb}
\usepackage{graphicx}
\begin{document}

\newcommand\zd{\mathcal{Z}}
\newcommand\hd{\mathcal{H}}   
\newcommand\met{\displaystyle{\not}E_T}

\preprint{PKU-ITP-091012}
\preprint{ANL-HEP-PR-09-96,~EFI-09-29}
\preprint{MSUHEP-091007}

\title{A Dark Matter Model with Non-Abelian Gauge Symmetry}
\author{Hao Zhang}
\email{haozhang.pku@pku.edu.cn}
\author{Chong Sheng Li}
\email{csli@pku.edu.cn}
\affiliation{Department of Physics and State Key Laboratory of Nuclear Physics
and Technology, Peking University, Beijing 100871, China}

\author{Qing-Hong Cao}
\email{caoq@hep.anl.gov}
\affiliation{High Energy Physics Divison, Argonne National Laboratory, 
Argonne, Illinois 60439, U.S.A.}
\affiliation{Enrico Fermi Institute, University of Chicago, Chicago, Illinois
60637, U.S.A.}

\author{Zhao Li}
\email{zhaoli@pa.msu.edu} 
\affiliation{Department of Physics and State Key Laboratory of Nuclear Physics
and Technology, Peking University, Beijing 100871, China}
\affiliation{Department of Physics and Astronomy, Michigan State University, 
East Lansing, Michigan 48824, U.S.A}

\pacs{95.35.+d,~12.60.-i,~13.90.+i} 

\begin{abstract}
We propose a dark matter model in which the dark sector is gauged
under a new $SU(2)$ group. The dark sector consists of $SU(2)$ dark 
gauge fields, two triplet dark Higgs fields, and two dark fermion doublets 
(dark matter candidates in this model). The dark sector interacts with the SM
sector through kinetic and mass mixing operators. 
The model explains both PAMELA and Fermi LAT data very well and also satisfies 
constraints from both the DM relic density and Standard Model precision 
observables. The phenomenology of the model at the LHC
is also explored. 
\end{abstract}

\maketitle

\section{Introduction}

Despite that no evidence of new physics signal has been observed at the colliders yet, the observations from
cosmology reveal that more than 20$\%$ of the whole Universe is made up by the so-called
dark matter (DM)~\cite{Dunkley:2008ie}. But the Standard Model (SM) does not provide any candidate for the dark matter, 
so discovering dark matter would be an undoubtable evidence of new physics beyond the SM. Especially, a series
of cosmic ray and gamma-ray observations from INTEGRAL~\cite{Strong:2005zx}, ATIC~\cite{Chang:2008zzr}, PAMELA~\cite{Adriani:2008zr,Adriani:2008zq}
and Fermi LAT~\cite{Abdo:2009zk} have attracted extensively attention ~\cite{Finkbeiner:2007kk,Bergstrom:2008gr,Cirelli:2008jk,Barger:2008su,Huh:2008vj,ArkaniHamed:2008qn,Yin:2008bs,Ishiwata:2008cv,Ibarra:2008jk,Hooper:2008kv,Cao:2009yy,Meade:2009rb,Hooper:2009fj,Cheung:2009qd,Cheung:2009si,Bi:2009uj,Chen:2009ew,Ishiwata:2009pt,Zant:2009sv,Barger:2009yt,Meade:2009iu,He:2009ra}, which indicate that:
\begin{itemize}
\item INTEGRAL detected a gamma-ray pick at $511~\rm{keV}$ from the center of the galaxy, which can be explained with $e^+e^-$ annihilation there.
\item ATIC observed-electron positron excess from $300~\rm{GeV}$ to about $800\rm{GeV}$, while the
observation from PAMELA shows that there is a positron excess at $10-100~\rm{GeV}$.
\item Even though the bump-like structure observed by ATIC is not confirmed by Fermi
LAT, an excess between 200GeV and 1TeV still remains in the $e^++e^-$ spectrum.
\item DAMA/LIBRA and DAMA/NaI experiments reported positive results from the direct detection of dark
matter~\cite{Bernabei:2008yi}.
\end{itemize}
Apparently, it is very exciting to interpret the anomalous signals as being induced by the dark matter. Toward this 
end, key questions that need to be answered are the following:
\begin{itemize}
\item Why is there no hadronic anomaly detected in the cosmic ray experiments?
\item Where does the large boost factor ($\sim10^2 - 10^4$), which is necessary to interpret the
positron anomaly as the dark matter signal, originate from?
\item What generates the pick near the electron mass pole observed by INTEGRAL?
\item Why other direct search experiments do not observe any signal of dark matter?
\end{itemize}
All those mysterious questions might be related to different aspects of dark matter. Many
dark matter models have been proposed to address on one or more questions listed above,
but we expect there is a dark matter model which can explain all above experiments very well.
In this work we propose a dark matter model in which the dark sector is gauged
under a new $SU(2)$ group. The dark sector consists of $SU(2)$ dark 
gauge fields, two triplets dark Higgs fields, and two dark fermion doublets 
(dark matter candidates in this model). The SM sector interacts with the dark
sector through kinetic and mass mixing operators. Using two component 
dark matters, the model could explain both PAMELA and Fermi LAT data very well,  
while satisfies constraints from both the DM relic density and SM precision 
observables.

The paper is organized as follows. In Sec. II, we present the field content and the Lagrangian of our model. In Sec. III, we discuss the 
astrophysics and cosmology observations. In Sec. IV, we show some constraints from the SM precision observables. 
A global fit to the experimental data is given in Sec. V. In Sec. VI, we analysis the signals of this 
model at the LHC. Discussions and conclusions are given in Sec. VII. The anomalous dimension and the evolution of the relevant
operator, and  the decay properties of some new particles are shown in Appendix A and B, respectively.

\section{The dark matter model}

\begin{figure}[]
\includegraphics[width=0.8\textwidth]{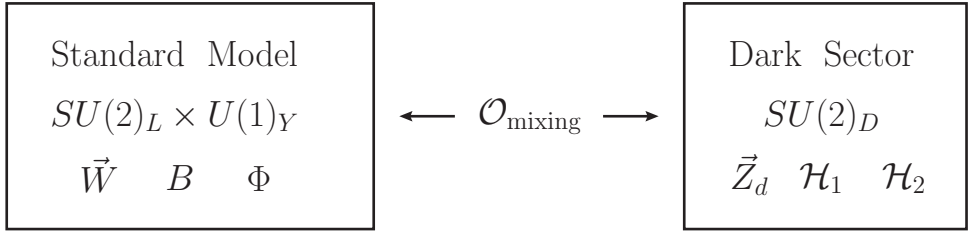}
\caption{Sketch of our non-Abelian dark matter model}
\label{fig:model}
\end{figure} 

In this section, we will propose a specific dark matter model. To answer the questions mentioned 
above, we adopt some ideas in references ~\cite{ArkaniHamed:2008qn,Finkbeiner:2007kk,TuckerSmith:2001hy}.
Now we present a non-Abelian dark matter model (NADM) in details. 
As shown in Fig.~\ref{fig:model}, our model consists of two sectors: one is
the usual SM sector, the other is the dark sector (DS). 
The SM sector is gauged under the usual $SU(3)_c\times SU(2)_L \times U(1)_Y$ symmetry 
while the dark sector under a new symmetry $SU(2)_D$ (named as dark symmetry
throughout this work). Particles in the dark sector, 
including the dark matter candidates, neither carry any SM 
charge nor interact with SM fields directly.
Similarly, all the SM particles are invariant under the dark gauge 
group transformation. In other words, both sectors are invisible to each other, 
but they can communicate via (i) the mixing between the SM Higgs boson and dark
scalar and (ii) high dimension operators induced by unknown ultraviolet (UV)
complete theory at a much higher energy scale $\Lambda$. 
The Lagrangian of our models is 
\begin{equation}
  \mathcal{L}
 =\mathcal{L}_{\rm SM} + \mathcal{L}_{\rm DS} + \mathcal{L}_{\rm mix},
\label{eq:DSM}   
\end{equation} 
where $\mathcal{L}_{\rm SM}$ ($\mathcal{L}_{\rm DS}$) denotes
the Lagrangian of the SM (dark) sector, and 
$\mathcal{L}_{\rm mix}$ represents the interaction between the SM and
dark sectors. Here the dark matter candidates will annihilate into
the dark gauge bosons and the dark scalar 
bosons, which eventually decay into the SM particles  
via mixing operators.

\subsection{Dark sector} 

The dark matter candidates in our NADM are Dirac fermion doublets $\psi_1$ and $\psi_2$.
The most economic way to obtain anomaly free is to use the vector-like
dark fermion.
Two or more dark fermion doublets are needed in order to fulfill both
XDM~\cite{Finkbeiner:2007kk} and iDM~\cite{TuckerSmith:2001hy} scenarios, because
both scenarios require different mass splits between the ground state and
excited state of dark matter candidate.
For example, the mass split is about $1~\rm{MeV}$ in the XDM scenario
but about $100~\rm{keV}$ in the iDM scenario.
In earlier works, some authors suggest the dark matter fermion is dark gauge
triplet~\cite{ArkaniHamed:2008qn,Chen:2009dm,Chen:2009ab}, which does not
need two generation DM fermions.
 
It is worth emphasizing the possibility of multiple dark matter candidates.
Many dark-matter candidates have been suggested in various models beyond 
the SM, but a nearly universal implicit assumption is 
that only one such candidate is needed and its properties are
constrained accordingly. Of course, no fundamental principle requires there is only one dark matter candidate, 
and the possibility of multipartite dark matter should not be 
ignored~\cite{Ma:2006uv,Hur:2007ur,Cao:2007fy,SungCheon:2008ts,Fairbairn:2008fb,Profumo:2009tb}.
Moreover, as to be shown later, more than one generation dark matters 
can fit all experimental data much better in our NADM.

The dark gauge symmetry could be broken in many ways, and one popular choice
is with the help of scalar field as the Higgs boson doublet in the SM. 
However, as pointed out in Ref.~\cite{Baumgart:2009tn}, the iDM scenario 
cannot be realized due to a custodial symmetry 
when only one dark scalar doublet is added. 
Rather than adding two complex scalar fields, 
we use two real scalar triplets (named as dark Higgs boson $\hd_i$) 
to break the custodial symmetry.
 
The Lagrangian of the dark sector is
\begin{eqnarray}
\mathcal{L}_{\rm{DS}}=&&
-\frac{1}{4}\mathcal{F}_{\mu \nu}^{a}\mathcal{F}^{\mu \nu a}+
\frac{1}{2}\sum_{j=1,2}\left(D_\mu \hd_j^a\right)
\left(D^\mu \hd_j^a\right)-V_{\hd}\left(\hd_1,\hd_2\right) \nonumber \\
&& + \bar{\psi_1}\left(i \displaystyle{\not}{\it{D}}-m_1\right)\psi_1
+\bar{\psi_2}\left(i \displaystyle{\not}{\it{D}}-m_2\right)\psi_2
+\sum_{j,k=1,2}\beta_{jk}\bar{\psi}_j \hd_k\psi_j, 
\label{eq:lagrangian_DS}
\end{eqnarray}
where the dark gauge field tensor $\mathcal{F}_{\mu\nu}^a$,
the dark Higgs field covariant derivative $D_\mu$, and the fermion
covariant derivative $\displaystyle{\not}{\it{D}}$ are defined by:
\begin{eqnarray} 
\mathcal{F}_{\mu\nu}^a &=& \partial_\mu \zd_{\nu}^a
-\partial_\nu \zd_{\mu}^a-g_D\epsilon^{abc}\zd_{\mu}^b \zd_{\nu}^c \\
D_\mu \hd_j^a&=&\partial_\mu \hd_j^a-g_D\epsilon^{abc}\zd_{d~\mu}^b \hd_j^c\\
\displaystyle{\not}{\it{D}}\psi_j&=&\gamma^\mu\left(\partial_\mu\psi_j
+ig_D\frac{\sigma^a}{2}\zd_{\mu}^a\psi_j\right).
\end{eqnarray}
Here, $g_D$ is dark gauge coupling constant, $V_{\hd}$ is the dark scalar
potential, $\beta_{jk}$ is the Yukawa coupling between dark Higgs boson 
and dark fermions, and $m_i$ is the intrinsic DM fermion mass for $\psi_i$.

We indicate that after symmetry breaking, both dark gauge bosons $\zd$ and dark fermions
$\psi_i$ obtain masses from the dark scalar vacuum condensation, and $m_{\zd}$ is
proportional to $g_D v_D$ while $m_i$ to $\beta_{ij} v_D$. 
As to be discussed below, a large mass hierarchy between $\zd$ and $\psi_i$,
say $m_{\zd} \ll m_i$, is required to fit the PAMELA data. Hence,
either a very light $\zd$ or a much heavy $\psi_i$ is preferred. Unfortunately,
the former requires a very small dark gauge coupling $g_D$ which will 
over produce DM relic abundance, and the latter demands a huge Yukawa coupling
which will destroy the vacuum stability. In order to avoid these problems, 
we explicitly keep the intrinsic fermion mass term $m_i$ in
$\mathcal{L}_{\rm DS}$. 
Such large masses could be generated by other exotic heavy fields 
which decouple at the scale much higher than the electroweak scale. 
The Yukawa interaction $\beta_{jk}\bar{\psi}_j \hd_k\psi_j$ would
generates mass split between the two components inside one fermion doublet.

Last, we comment on the dark scalar potential. When two and more scalar 
multiplets present, e.g. as in two Higgs doublet model, the scalar potential 
is rather complicated. Instead of discussing a general scalar potential, 
we will choose the scalar potential in Eq.~\ref{eq:lagrangian_DS} as a simple form, 
\begin{equation}
V_{\hd}\left(\hd_1,\hd_2\right)=\frac{1}{4}\lambda
\left(  \sum_{j=1,2}
\left(\hd_j^a \hd_j^a-v_d^2/2\right)^2 -2\left(\hd_1^a \hd_2^a \right)^2
\right),      
\end{equation}
which preserves a stable vacuum and breaks the custodial symmetry
simultaneously. Here $v_d$ denotes the vacuum expectation value (VEV) of 
dark Higgs bosons.

\subsection{Interaction between the SM and dark sectors}  

\begin{figure}[]
\includegraphics[width=0.8\textwidth]{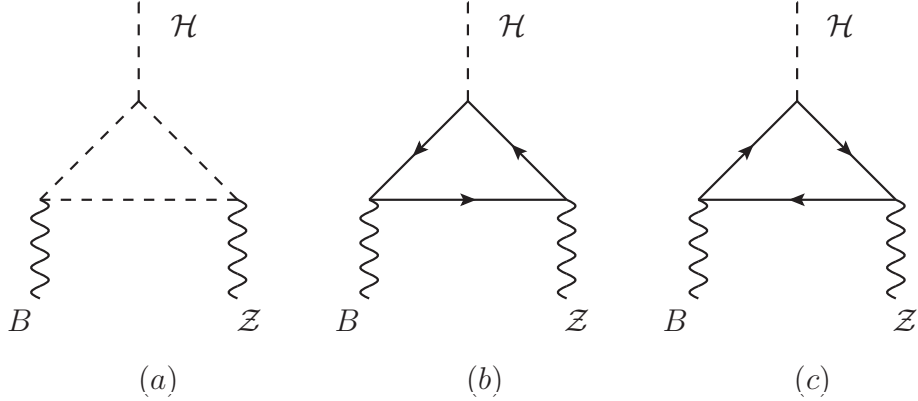}
\caption{Representative Feynman diagrams which can induce the dimension-five
operator: (a) heavy boson loop, (b) and (c) heavy fermion loop.}
\label{fig:eff_op}
\end{figure}

We first consider that the dark sector interacts with the SM sector through 
a dimension-5 operator 
\begin{equation}
\mathcal{L}_{\rm mix} \supset
-\frac{1}{2\Lambda} B^{\mu\nu}
\left(\alpha_1 \hd_1^a+\alpha_2 \hd_2^a\right) \zd_{\mu\nu}^{a},
\label{eq:mixoperator}      
\end{equation}
where $B^{\mu\nu}$ is the field strength tensor of the gauge boson associated
with the SM $U(1)_Y$ group and $\Lambda$ is the cutoff scale
of new physics (NP). 
Such an operator could be induced by integrating out new heavy particles
in a renormalizable theory as shown in Fig.~\ref{fig:eff_op}, which could generate the dimension-5 operator when new heavy
particles decouple. The evolution behavior of this operator is determined by 
its anomalous dimension
(see Appendix A). 

Another interaction between the SM and dark sectors is via the mixing between
the dark scalar and the SM Higgs boson, such as
\begin{equation}
\mathcal{L}_{\rm mix} \supset \epsilon_{\hd}
(\Phi^{\dagger}\Phi)(\hd_{1}\cdot \hd_{2}),
\label{eq:scalarmixing}   
\end{equation}  
where $\Phi$ denotes the SM Higgs doublet and $\epsilon_{\hd}$ is the mixing
parameter. Note that the scalar mixing is crucial to prevent overproducing
DM relic abundance in our model.  For example, without the scalar mixing, 
there exists a stable dark scalar after symmetry breaking, which could also
be dark matter candidate as long as it is lighter than dark fermions. 
Even though the dark scalar pair could annihilate into the SM particles through
the process $\hd \hd \to \gamma \gamma \to {\rm SM~particles}$,
the annihilation cross section is too small to produce correct amount of relic
abundance.  
In fact, it produces too much relic abundance which would overclose 
the Universe. Such a problem is solved by the scalar mixing with
the SM Higgs boson, which enables 
the lightest dark scalar decaying into SM fermion.

Next, we discuss some theoretical constraints on $\epsilon_{\hd}$:
\begin{itemize}
\item {\it Vacuum stability }: 
The existence of the lower bound of the scalar potential requires 
$$\epsilon_{\hd}v_{\rm SM}v_d/m_{H}<M_{\mathcal{H}},$$
where $v_{\rm SM}$ and $m_H$ are the vacuum exception value and the mass of SM Higgs boson ($H$), respectively.
\item {\it Naturalness }:
As pointed out in Ref.~\cite{Finkbeiner:2008gw}, the naturalness condition
requires that $\epsilon_\hd v_{\rm SM}^2$ is not much larger than $v_d^2$,
i.e. $$\epsilon_\hd v_{\rm SM}^2\sim v_d^2,$$
which leads to the following bounds:
(1) $\epsilon_\hd \sim 10^{-5}$ for $v_d \sim 1~\rm{GeV}$ and
(2) $\epsilon_\hd \sim 10^{-7}$ for $v_d \sim 0.1~\rm{GeV}$.
\end{itemize}

\subsection{Symmetry breaking and mass spectrum of dark particles}

In our model the dark $SU(2)_D$ gauge symmetry is broken spontaneously
when the two dark scalar triplets develop non-zero vacuum expectation values (VEV).
The dark scalar potential achieves its minimum value at
\begin{equation}
\left< \hd_{\rm 1} \right>=\frac{1}{\sqrt2}
\begin{pmatrix}
~0~\\
~0~\\
~v_d~
\end{pmatrix}
~~\rm{and}~~
\left< \it{\hd}_{\rm{2}} \right>=\frac{1}{\sqrt2}
\begin{pmatrix}
~0~\\
~v_d~\\
~0~
\end{pmatrix},
\end{equation}
where, for simplicity, we choose same VEV ($v_d$) for both scalar triplets. 
Therefore, the non-Abelian $SU(2)_D$ symmetry is totally broken by the non-zero $v_D$.
 
After dark gauge symmetry breaking,  all the three dark gauge bosons become massive:
\begin{equation}
   m_{\zd_2} = m_{\zd_3} = M_{\mathcal{Z}} = g_D v_d,\quad {\rm and}\quad
   m_{\zd_1} = \sqrt 2 M_{\mathcal{Z}}. 
\end{equation}
The dark gauge boson mass will be modified slightly when they mix
with SM gauge bosons through the operator as shown in
Eq.~\ref{eq:mixoperator}. Without loosing generalization, 
we choose $\alpha_1=0$ in Eq.~\ref{eq:mixoperator}. 
As a result, only $\zd_2$ mixes with SM gauge boson.  
Note that a non-zero $\alpha_1$ merely changes the definition of 
mass eigenstates, and there is still only one dark gauge boson 
mixing with SM gauge boson directly. 
Then, the kinematic mixing between $U(1)_Y$ gauge boson and
$\zd_2$ generated by this operator is
\begin{equation}
\mathcal{L}_{KM}=-\frac{\epsilon}{2} B_{\mu\nu} \zd_2^{\mu\nu},
\end{equation}
where $\epsilon =\alpha_2 v_d /\sqrt{2}\Lambda$, and $\zd_2^{\mu\nu}=\partial^{\mu}\zd_2^{\nu}-\partial^{\nu}\zd_2^{\mu}$. 
It leads to non-diagonal elements in the kinetic energy term of Lagrangian in the basis
$\hat{V}^T_{\mu} = (\hat{\zd}_{2\mu}, \hat{B}_\mu, \hat{W}_{3\mu})$
\begin{equation}
 \mathcal{K} =
\begin{pmatrix}
1        & \epsilon & 0 \\
\epsilon & 1        & 0 \\
0        & 0        & 1
\end{pmatrix}
\label{eq:KE_matrix},
\end{equation}
where the gauge field with a caret symbol is understood as the current eigenstate.
Furthermore, the mass matrix takes the following form 
\begin{equation}
 \mathcal{M}^2 =
\begin{pmatrix}
M_{\mathcal{Z}}^2 &            0                 &          0           \\
0   & \frac{1}{4}v_{SM}^2 g_Y^2    & -\frac{1}{4} g_2 g_Y \\
0   & -\frac{1}{4}v_{SM}^2 g_2 g_Y & \frac{1}{4} v_{SM}^2g_2^2 
\end{pmatrix}
\label{eq:Mass_matrix}
\end{equation}
We denote the mass eigenstates of the three gauge bosons by 
$V^T_{\mu} = (\zd_{2\mu}, A_\mu, Z_{\mu})$. A simultaneous diagonalization of both the kinetic energy term 
and the mass matrix gives (to order $\epsilon$)
\begin{eqnarray}
&&\hat{B}^\mu=A^\mu c_W-Z^\mu s_W+\left(\frac{\tilde{m}_Z^2 s_W^2}{\tilde{m}_Z^2-M_{\mathcal{Z}}^2}-1\right)\epsilon \zd_2^\mu,\nonumber \\
&&\hat{W}_3^\mu=A^\mu s_W+Z^\mu c_W-\frac{\tilde{m}_Z^2 s_W c_W\epsilon}{\tilde{m}_Z^2-M_{\mathcal{Z}}^2} \zd_2^\mu,\nonumber \\
&&\hat{\zd}_2^\mu=\zd_2^\mu+\frac{\tilde{m}_Z^2 s_W\epsilon}{\tilde{m}_Z^2-M_{\mathcal{Z}}^2} Z^\mu,
\end{eqnarray}
where $s_W \equiv \sin\theta_W = g_Y/\sqrt{g_2^2+g_Y^2}$,  
$c_W \equiv \cos\theta_W = g_2/\sqrt{g_2^2+g_Y^2}$. 
And the masses of the corresponding mass eigenstates are given by (to order of $\epsilon^2$)
\begin{eqnarray}
& & m_A^2=0,\nonumber \\
& & m_Z^2=\tilde{m}_Z^2\left(1+\frac{\tilde{m}_Z^2 s_W^2}
{\tilde{m}_Z^2-M_{\mathcal{Z}}^2}\epsilon^2\right), \nonumber \\
& & m_{\zd_2}^2=M_{\mathcal{Z}}^2\left(1-\frac{\tilde{m}_Z^2 s_W^2}
{\tilde{m}_Z^2-M_{\mathcal{Z}}^2}\epsilon^2\right),
\label{eq:masses}   
\end{eqnarray}
where $\tilde{m}_Z$(=$g_2 v_{SM}/2 c_W$) is the SM $Z$-boson mass.
The gauge fields in the mass eigenstate basis are (to order of $\epsilon$) 
\begin{eqnarray}
&&A^\mu=c_W \hat{B}^\mu+s_W \hat{W}_{3}^\mu -\epsilon \hat{\zd}_2^\mu c_W,
\nonumber \\
&&Z^\mu=c_W \hat{W}_{3}^{\mu} - s_W \hat{B}^{\mu}
+\epsilon \frac{M_{\mathcal{Z}}^2 s_W}{\tilde{m}_Z^2-M_{\mathcal{Z}}^2} \hat{\zd}_2^\mu,
\nonumber \\
&&\zd_2^\mu=\hat{\zd}_2^{\mu}
+\epsilon\frac{\tilde{m}_Z^2 s_W}{\tilde{m}_Z^2-M_{\mathcal{Z}}^2} (c_W \hat{W}_{3}^{\mu} - s_W \hat{B}^{\mu}).
\end{eqnarray}

Two dark scalar triplets carry six degrees of freedom. While three of them are
eaten by dark gauge boson after symmetry breaking, the rest three remain as 
three physical dark scalar fields.  In the unitary gauge, the two scalar triplets can be written as
\begin{equation}
\hd_{1}=
\left(
\begin{array}{c}
   0          \\
   h_{3}/\sqrt{2}      \\
   h_{1}+v_d/\sqrt{2}
\end{array}
\right)
~~{\rm and}~~
\hd_{2}=
\left(
\begin{array}{c}
   0          \\
   h_{2}+v_d/\sqrt{2}  \\
   h_{3}/\sqrt{2}
\end{array}
\right),
\end{equation}
where $h_1$, $h_2$ and $h_3$ are physical dark scalars with degenerate mass
$M_{\mathcal{H}}^2=\lambda_D v_d^2=\lambda_D M_{\mathcal{Z}}^2/g_D^2$. 
In addition the mixing term in Eq.~\ref{eq:scalarmixing} can induce the following mass matrix in the basis ($H_0$, $h_3$)
\begin{equation}
\frac{1}{2}
\left( \begin{array}{cc} H_{0} & h_{3}\end{array}\right)
\left(\begin{array}{cc}
m_H^2                  & \epsilon_{\hd} v_d v_{SM} \\
\epsilon_{\hd}v_d v_{SM} & M_{\mathcal{H}}^2
\end{array}\right)
\left(\begin{array}{c} 
H_{0}\\
h_{3}\end{array}\right).
\end{equation}
Thus, the mass eigenstates and their masses are given by
\begin{eqnarray}
\tilde{h}_3 = h_{3}-\frac{\epsilon_{\hd}v_dv_{SM}}{m_{H}^{2}-M_{\mathcal{H}}^{2}}H_{0}
&,&
m_{\tilde{h}_3}^2 = 
M_{\mathcal{H}}^{2}-\frac{\epsilon_{\hd}^{2}v_d^{2} v_{SM}^{2}}{m_{H}^{2}-M_{\mathcal{H}}^{2}} 
\nonumber \\ 
\tilde{H}_0 = H_{0}+\frac{\epsilon_{\hd}v_d v_{SM}}{m_{H}^{2}-M_{\mathcal{H}}^{2}}h_{3}
&,&
m_{\tilde{H}_0}^2 =
m_{H}^{2}+\frac{\epsilon_{\hd}^{2}v_d^{2} v_{SM}^{2}}{m_{H}^{2}-M_{\mathcal{H}}^{2}}. 
\end{eqnarray}      
The dark Higgs boson $\tilde{h}_3$ is no longer stable as it can decay to 
SM fermion pair through the above mixing.

\section{Astrophysics and Cosmology Observations}
\subsection{Dark matter relic density}
In our model the dominant contribution to dark matter annihilation comes from 
the channels involving dark gauge boson and dark Higgs (also with the lighter 
generation dark fermion for the heavier generation freezing out) in the final 
state, which is of order $\alpha_D^2$. 
Those annihilation products decay into the SM particles eventually. Below we
calculate the relic abundance in our model following Ref.~\cite{Kolb:1990}.  
Furthermore, coannihilation effects~\cite{Griest:1990kh} are also considered 
in our calculation, as the masses of the two component fields of DM fermion 
doublet are nearly degenerated. 

As to be shown later, in order to fit both PAMELA and Fermi LAT data in our 
model, it is necessary to have a large mass splitting between $\psi_1$ and 
$\psi_2$. Therefore, the DM annihilation involves the dark gauge interaction at 
different energy scales. Running effect of dark gauge coupling has to be
considered in the numerical evaluation.
The running behavior of dark gauge coupling is governed by renormalization
group equation (RGE). The one-loop level beta function for a general 
non-Abelian gauge theory is~\cite{Gross:1973ju}
\begin{equation}
\beta\left(g\right)=\frac{g^3}{16\pi^2}\left(-\frac{11}{3}N+\frac{1}{6}C_{2b}+\frac{4}{3}C_{2f}\right).
\end{equation}
In our model, $N=2$, $C_{2b}=2n_b$ and $C_{2f}=\frac{n_f}{2}$, where $n_b=2$ is 
the number of real Higgs triplet, $n_f$ is the number of fermion doublet. 
For the energy scale $\mu>m_2$, $n_f=2$, for $m_1<\mu<m_2$, 
$n_f=1$, and for $\mu<m_1$, $n_f=0$. Within this theoretical framework, 
the dark gauge interaction is asymptotic free, i.e. becoming stronger at low
energy but weaker at high energy. 
In this work we choose the following parameters as benchmark:
\begin{eqnarray}
m_1=510~\rm{GeV} &,& m_2=1300~\rm{GeV},\nonumber \\
M_{\zd} = 0.1~\rm{GeV} &,& M_{\hd} = 0.02~\rm{GeV},
\end{eqnarray} 
to explore the impact of RG running effect on the relic abundance.

\begin{figure}[ht!]
\includegraphics[width=0.9\textwidth]{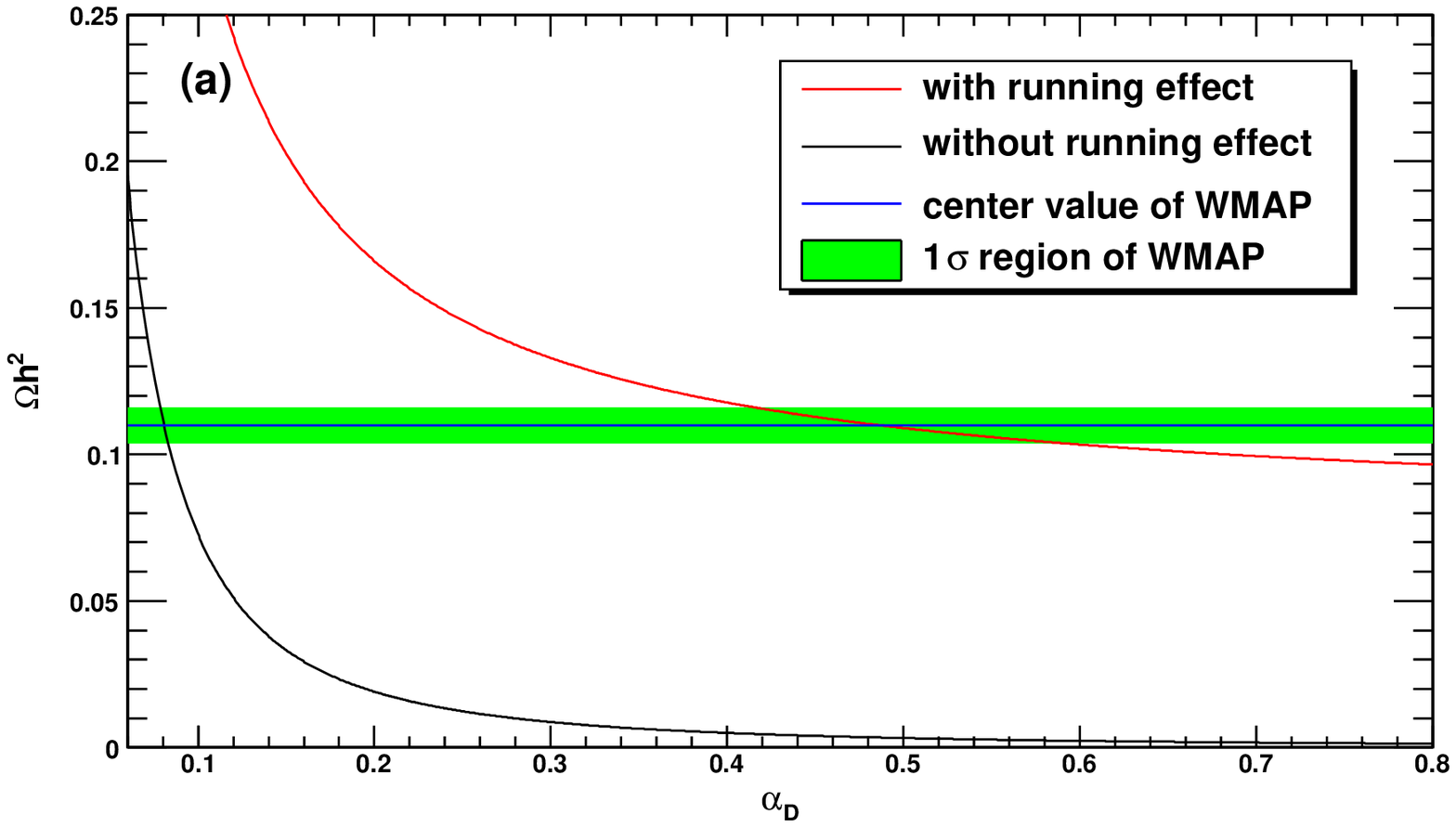}\\
\includegraphics[width=0.9\textwidth]{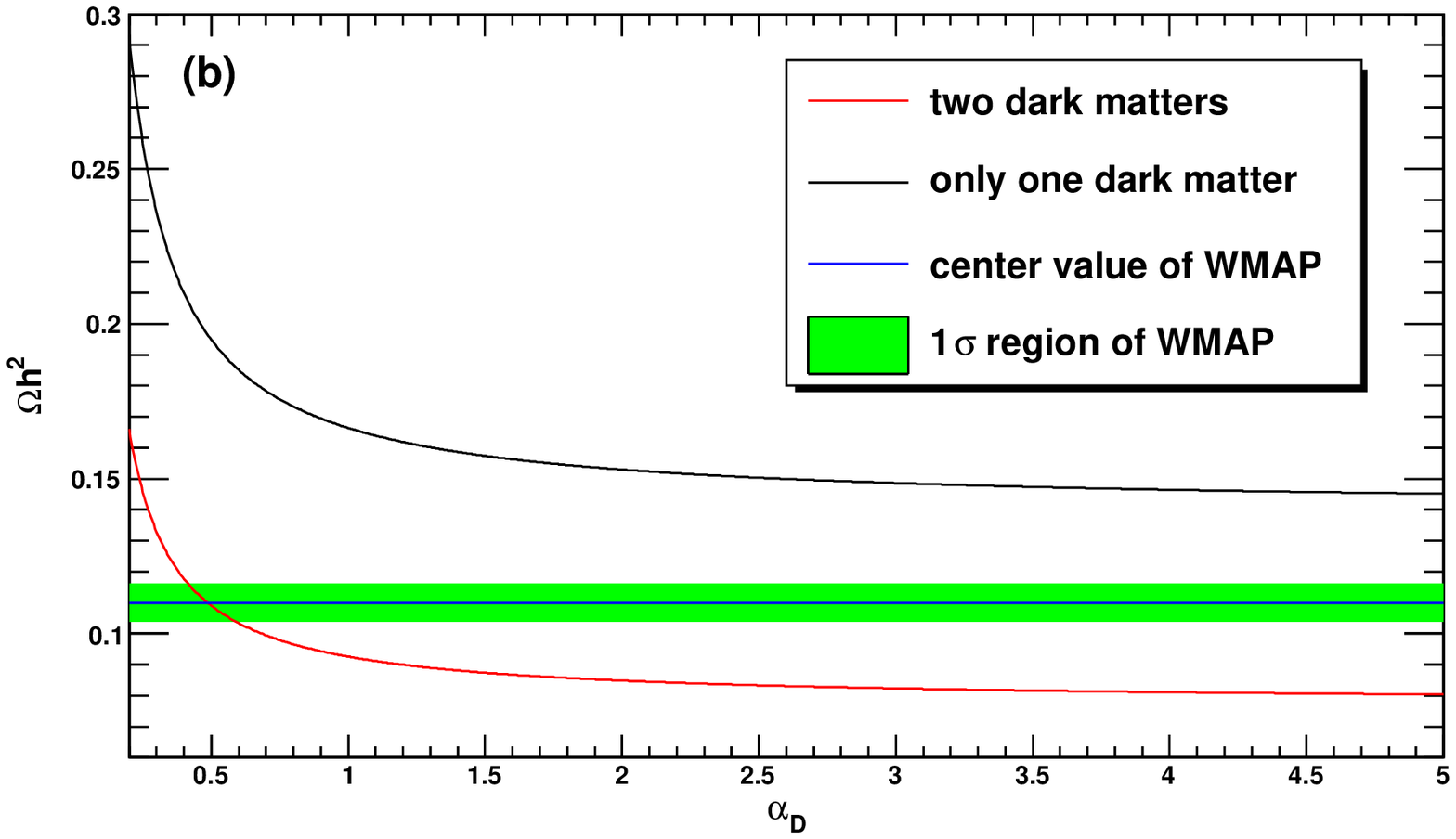} 
\caption{
(a) Relic abundance versus dark gauge coupling evaluated at the scale 
$M_{\mathcal{Z}}$, with and without running effect of dark gauge boson. 
The annihilation cross sections are calculated with the dark gauge coupling 
at the scale $\mu=2m_i$. 
The parameters are chosen to be $m_1=510~\rm{GeV}$, $m_2=1300~\rm{GeV}$,
$M_\zd=0.1~\rm{GeV}$ and $M_{\hd}=0.02~\rm{GeV}$. 
The dark green band display the cold dark matter density from 
Five-Year WMAP data, $\Omega h^2=0.1099 \pm 0.0062$~\cite{Dunkley:2008ie}.
(b) Comparison of two generation DMs and one generation DM with the choice
of above parameters, except that we fix the DM mass be $1300~\rm{GeV}$ in 
the one dark matter case. } 
\label{fig:relic}
\end{figure}

Figure~\ref{fig:relic}(a) displays the relic density versus dark gauge boson 
coupling at the scale $\mu=M_{\zd}$. 
Note that, with running effects, the dark gauge coupling is evaluated at the
scale $\mu=2m_i$ in the numerical calculation of DM annihilation 
cross sections. 
It is well known that the relic density of DM is approximately given
by $\Omega h^2 \approx (0.1~\rm{pb})/\left<\sigma v\right>$ where 
$\left< \sigma v\right>$ is the thermally averaged product of the DM
annihilation cross section with its velocity. Five-year WMAP data indicates
$\left< \sigma v\right> \simeq \rm{pb}$. Since the final state particles in
the DM annihilation are almost massless, the annihilation cross section can be
written as following, based on the dimension counting,
\begin{equation}
\sigma_i v \propto \frac{g_D^2}{m_i^2}.  \label{eq:relic}    
\end{equation}  
When the DM candidate masses are given, there exists an unique dark gauge 
coupling to produce the correct amount of relic abundance. For example, 
without running effect, the correct relic density can be produced only 
in the vicinity of $\alpha_D \sim 0.1$; see the black curve in 
Fig.~\ref{fig:relic}(a). With running effects, the gauge coupling at the scale
$\mu = 2 m_i$ is less than its value at the scale $\mu=M_\zd$, 
therefore, large coupling strength is needed to induce efficient 
DM annihilation; see the red curve in Fig.~\ref{fig:relic}(a). Owing to the RG
running effect, the favored coupling region, which is consistent with the WMAP
data at $1\sigma$ level (green band), become much broader (e.g. 
$0.45<\alpha_D<0.67$) for the given benchmark parameters. 

In Fig.~\ref{fig:relic}(b), we display the comparison of two dark matters (red) 
and  one dark matter (black). In the case of one generation dark matter, 
we fix its mass be 1300~GeV because the Fermi LAT data prefers a very heavy
DM candidate. 
It is clear that the model cannot produce a correct DM relic density when only 
one heavy DM candidate presents, even for an unreasonably large coupling such 
as $\alpha_D\sim 5$. It is due to the heavy mass suppression in 
Eq.~\ref{eq:relic} which yields a small DM annihilation rate, 
cf. the black curve. Introducing another generation dark matter efficiently 
reduces relic density to satisfy the WMAP data, cf. the red curve.

\subsection{INTEGRAL and XDM}
As pointed out in Ref.~\cite{Finkbeiner:2007kk}, if dark matter has an excited 
state and the mass split is about $2 m_e$, decay of the excited state of 
dark matter may produce much $e^+e^-$ as a source of the gamma-ray line 
in the galactic center at $510.954\pm0.075~\rm{keV}$ confirmed by
INTEGRAL/SPI~\cite{Churazov:2004as}. 
Following the proposal in Ref.~\cite{Finkbeiner:2007kk}, we find that  
in our model the dark gauge boson mass $M_{\mathcal{Z}}$ should be less than
0.2~GeV in order to give a significant contribution to the 511~keV signal.  
As a result, such a light dark gauge boson cannot decay into quarks,  
muon and tau due to the kinematics.

\subsection{PAMELA and Fermi LAT}
To fit the data of PAMELA and Fermi LAT, we calculate the cosmic ray $e^+e^-$ 
flux,  using the background flux formula~\cite{Baltz:1998xv} and the formulas
of the dark-matter-induced flux~\cite{Cirelli:2008id}. 
In Appendix B we summarize the decay properties of dark gauge boson and dark 
scalars used in our numerical calculation. 
The large Sommerfeld enhancement factor, which is given by the non-Abelian dark 
gauge interaction, can be calculated numerically following Ref.~\cite{Iengo:2009xf}. 
The results of fitting PAMELA and Fermi LAT is shown below.

\subsection{Direct detection of dark matter}
Current direct detection experiments of dark matter ~\cite{Alner:2007ja,Bernabei:2008yi,Angle:2007uj,Ahmed:2008eu} give negative result except DAMA~\cite{Bernabei:2008yi}.  In our model, the interaction between  
dark matters and nucleon is iDM like~\cite{TuckerSmith:2001hy}, so the direct detection can not give a strong limit 
to $\epsilon$. For the generation whose mass split is greater than $2m_e$ required by XDM, the corresponding 
critical velocity~\cite{TuckerSmith:2001hy} will be larger than the escape velocity of dark matter in the galaxy, which means that it can not be detected in the direct detection experiments. For another generation, the mass split is still a free 
parameter which can be use to explain DAMA and other direct detection experiments data. 

\section{constraints from the SM precision observables}
Due to its mixing with the SM gauge bosons, the dark gauge boson inevitably
modifies various low energy precision observables, such as the oblique 
parameters~\cite{Peskin:1991sw}, muon anomalous magnetic momentum, 
and $Z^0$ boson decay, etc. In this section we explore
constraints from those precision measurements.

\paragraph{The Electroweak Oblique Parameters} 
The oblique parameters $S$, $T$ and $U$ can be calculated with formulas
in Ref.~\cite{Babu:1997st}. In our model, to the order of $\epsilon^2$, 
\begin{eqnarray}
&&S=\frac{2\sqrt{2}\pi\epsilon^2}{G_F\left(m_Z^2-M_{\mathcal{Z}}^2\right)},\nonumber\\
&&T=\frac{\pi\epsilon^2}{\sqrt{2}c_WG_F\left(m_Z^2-M_{\mathcal{Z}}^2\right)},\nonumber\\
&&U=0,
\end{eqnarray}
where $G_F$ is the Fermi constant.
For $M_{\zd}\sim 0.1~\rm{GeV}$, from above parameters we can get 
$\epsilon \lesssim 0.023$ at $1\sigma$ level.

\paragraph{Muon Anomalous Magnetic Moment}
The additional contribution to the muon anomalous magnetic moment 
$\left(g-2\right)_{\mu}$ is mainly from $\zd_2$, which can be calculated by ~\cite{Bars:1972pe}
\begin{equation}
F_2(0)=\frac{2g_2^2}{m_{\zd_2}^2}m_{\mu}^2\int_0^1dz(1-z)\int\frac{d^4k}{(2\pi)^4}\frac{1}{(k^2+b^2)^3}\left(h_1(z)+\frac{g_1^2}{g_2^2}h_2(z)\right),
\end{equation}
with
\begin{eqnarray}
&&h_1(z)=-4z(z+3),\\
&&h_2(z)=4z(1-z),\\
&&b^2=z+\frac{m_{\mu}^2}{m_{\zd_2}^2}(1-z)^2,
\end{eqnarray}
and
\begin{eqnarray}
&&g_1^2=\frac{e^2\epsilon^2}{16c_W^2}\left[4c_W^2+\frac{M_{\mathcal{Z}}^2}{M_Z^2}(1-4s_W^2)\right]^2, \nonumber\\
&&g_2^2=\frac{e^2\epsilon^2}{16c_W^2}\frac{M_{\mathcal{Z}}^4}{M_Z^4},
\end{eqnarray}
where $m_{\mu}$ is the muon mass.
For $\epsilon\lesssim7\times10^{-4}$, the contribution from $\zd_2$ to $\left(g-2\right)_{\mu}$ is less than 
$10^{-10}$. Here we choose $M_{\mathcal{Z}}\sim0.1$GeV and $M_{\mathcal{H}}\sim0.02$GeV. 

The contribution from $\tilde{h}_3$ is much less than $\zd_2$. Even for $\epsilon_{\hd}\sim400$ the contribution from $\tilde{h}_3$ to $\left(g-2\right)_{\mu}$ is less than 
$10^{-10}$ when $M_{\mathcal{Z}}\sim0.1$GeV and $M_{\mathcal{H}}\sim0.02$GeV. These results show that the constraint from $\left(g-2\right)_{\mu}$ is not strong. 

\paragraph{Decay width of the SM $Z^0$ Boson}
In our model there are some new decay channels of the SM $Z^0$: $Z^{0}\rightarrow \zd_{1}\zd_{3},~Z^{0}\rightarrow \zd_{2}h_{1},~Z^{0}\rightarrow \zd_{2}h_{2},~Z^{0}\rightarrow \zd_{2}\tilde{h}_{3},~Z^{0}\rightarrow \zd_{3}\tilde{h}_{3},~Z^{0}\rightarrow\gamma h_{2}$. 
Ignoring the higher order terms of $M_\hd^2/m_Z^2$ 
and $M_\zd^2/m_Z^2$, to the order of $\epsilon^2$, 
the non-zero decay widths are
\begin{eqnarray}
\Gamma^Z_{\zd_{1}\zd_{3}}=\Gamma^Z_{\zd_{2}h_{2}}=2\Gamma^Z_{\zd_{3}\tilde{h}_{3}}=\frac{\alpha_D \epsilon^2s_W^2 m_Z^3}{12 M_{\mathcal{Z}}^2},
\end{eqnarray}
\begin{eqnarray}
\Gamma^Z_{\zd_{2}h_{1}}=\frac{\alpha_{D}\epsilon^{2}s_{W}^{2}m_Z}{24}.
\end{eqnarray}
From above results and the experimental data of the $Z^0$ decay width ~\cite{Amsler:2008zzb}, at $1~\sigma$ level, 
$\epsilon$ should be less than $4.5 \times 10^{-5}$ for $M_\zd \sim 0.1$GeV, $\alpha_D(M_\zd)=0.57$ and $M_{\hd}\sim0.02$GeV. 
Our calculation shows that the evolution of the operator in Eq.~\ref{eq:mixoperator} is negligible small (see Appendix A).

\section{Explanation of the PAMELA and Fermi LAT observations}

In this section, we present a global fit of the PAMELA, Fermi LAT, DM relic density, the $S$ and $T$ parameters, and $Z^0$ boson decay width, using least $\chi^2$ analysis~\cite{Stump:2001gu,Li:2009br}. 
As widely used in the literatures, a parameter $b_k$ is introduced to rescale the cosmic ray background here.
In general, the total distribution of dark matter in the galaxy can be determined by N-body simulations, 
but for a model of two generation dark matters, the fraction of each generation in the galaxy is unknown.
In our model, since the masses of two dark matters are different, the ratio $r_{\rm{galaxy}}$ of their total masses in the galaxy will be different 
from $r_{\rm{freeze}}=\Omega_{\rm{heavy}}/\Omega_{\rm{light}}$, where $\Omega_{\rm{heavy}}(\Omega_{\rm{light}})$ is the relic density of the heavier (lighter) dark matter. So we take $r_{\rm{galaxy}}$ as a free parameter.

In our fitting we require the theoretical values of both the SM relevant precision observables and the relic density fall in 1$\sigma$ region of the experimental data.
We fit the PAMELA and Fermi LAT data using NFW profile~\cite{Navarro:1995iw} 
and ``MED'' propagation scheme~\cite{Cirelli:2008id}. 
The best-fit solution is 
$\alpha_D(M_\zd)=0.57$, $M_\zd=0.1~\rm{GeV}$, $M_\hd=0.02~\rm{GeV}$, 
$m_1 = 510~\rm{GeV}$, $m_2=1300~\rm{GeV}$, $\epsilon=10^{-5}$, $b_k=0.66$ 
and $r_{\rm{galaxy}}=6.1$, resulting in $\chi^2 = 1.1$ per degree of freedom.

\begin{figure}
\includegraphics[width=1.0\textwidth]{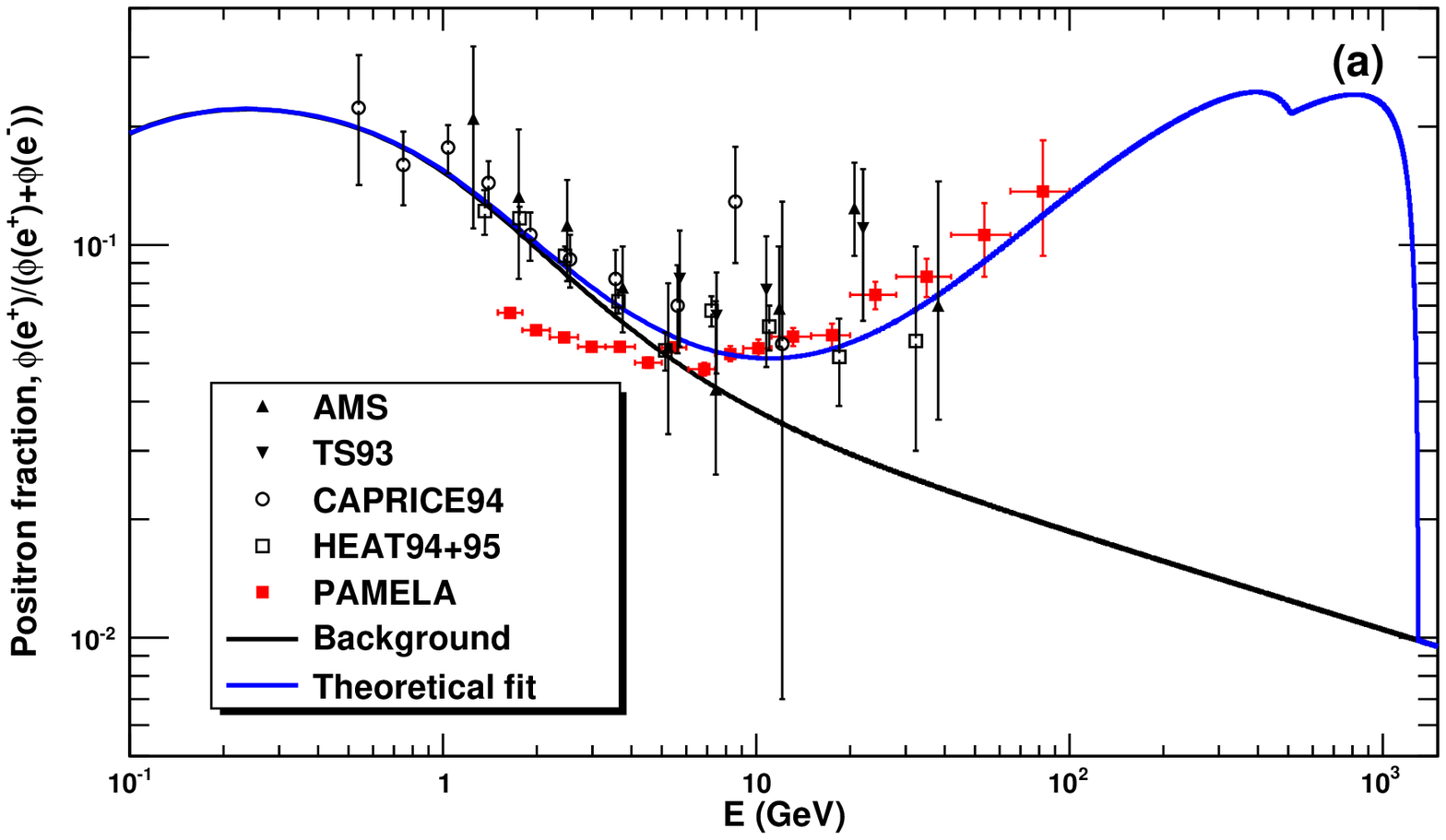}\\
\includegraphics[width=1.0\textwidth]{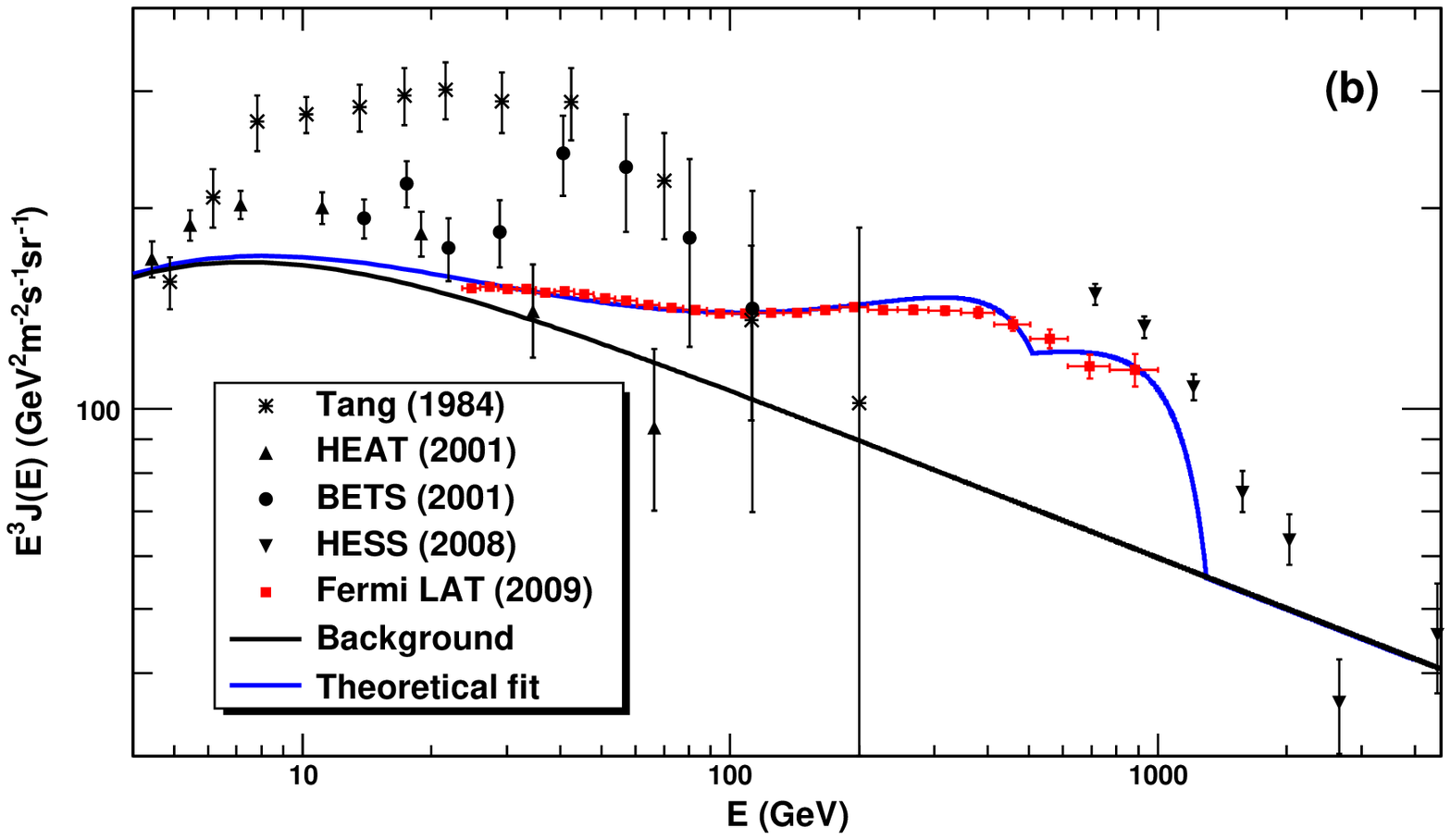}  
\caption{
(a) Fit to PAMELA data of this model at the minimum $\chi^2$ parameter 
point where the experiments data can be found in ~\cite{Adriani:2008zr,Aguilar:2007yf,Golden:1995sq,Boezio:2000ha,Barwick:1997ig}; 
(b) Fit to Fermi LAT data of this model at the minimum $\chi^2$ parameter point 
where the experiments data can be found in ~\cite{Abdo:2009zk,Tang:1984ev,DuVernois:2001bb,Torii:2001aw,Collaboration:2008aaa}. } \label{fig:pamela-fermi}
\end{figure}

We illustrate, in Fig.~\ref{fig:pamela-fermi}(a), the positron fraction as seen
at Earth after propagating effects are included in the ``MED'' propagation
scheme~\cite{Cirelli:2008id}. We note that our best fit is quite 
good, with the Sommerfeld enhancement boost factor of 
$\sim5\times10^3$ in our case. 
The kink at the high energy 
region is due to 
the overlap of cosmic positrons from two component DM annihilations. 
Figure~\ref{fig:pamela-fermi}(b) shows that 
our model prediction also fits the Fermi LAT very well.  
With enough precision, the kink feature of two 
component dark matters may be explored in future experiments.

\section{LHC phenomenology}

The typical signals of our model at the LHC are the lepton jets, which are boosted groups of $n\geqslant2$ leptons with small angle separations and GeV scale invariant masses as shown in Ref.~\cite{ArkaniHamed:2008qp}. 
Especially, for the process $p+p\rightarrow Z^0 \to \zd_2+h_2\rightarrow e^+e^-+\displaystyle{\not}E_T$, the signal is just one lepton jet originated
from the $\zd_2$ decay plus $\met$ from the long lived 
$h_2$. See Appendix~\ref{sec:decay} for detailed calculation of the dark gauge
boson and dark scalar decays. 
The intrinsic SM backgrounds originate from the $W^+W^-$ and $ZZ/Z\gamma^*$ 
pair production. Only the $W\to e \nu_e$ decay mode is considered in the $WW$ 
background in order to mimic the signal signature. Similarly one of the $Z$ 
boson has to decay into neutrino pair and the other $Z$ boson or off-shell 
photon decays into electron-positron pair. 
With the help of CalcHEP~\cite{Pukhov:2004ca}, we examine the kinematics 
distributions of both signal and background.  

\begin{figure}
\includegraphics[width=0.9\textwidth]{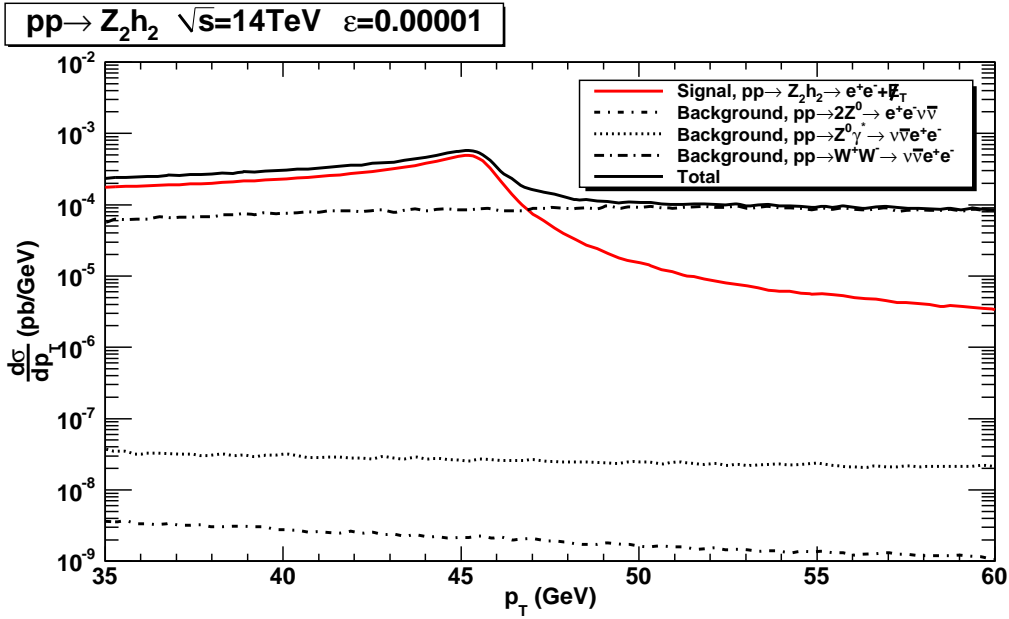}
\caption{
Differential distribution of transverse momentum of the lepton jet 
system $p_T^{e^+e^-}$ with the cuts: $m_{e^+e^-}<10$GeV and 
$\cos\theta_{e^+ e^-}>0.95$. } 
\label{fig:lhc}
\end{figure}

Kinematics of the signal process is very distinctive from those of background 
processes: 
\begin{itemize}
\item {\it Signal: }
The two charged leptons in the signal are forced to move
collaterally due to the large boost received from $\zd_2$. Even though it is
challenging to measure the momentum of each charged lepton, one can measure the 
sum of the energy and transverse momentum ($p_T$) of the two charged lepton
system.  The distribution of $p_T^{e^+ e^-}$ as well as $\met$ peaks around 
half of the $Z^0$ boson mass; see the red-dotted curve in Fig.~\ref{fig:lhc}. 
Similar to the $W\to \ell \nu$ decay, the transverse mass of the lepton-jet 
and missing transverse momentum $\met$ also exhibit a sharp
Jacobin peak around $m_{Z^0}$. On the other hand, the invariant mass of the 
leptons is around GeV.
Last, the $\zd_2$  boson is long lived with $\epsilon=10^{-5}$. 
It, when produced, will propagate about 30 centimeter inside the detector,
which provides an unique collider signature.

\item {\it $WW$ background: } 
The two charged leptons in the $WW$ background neither move parallel
to each other nor exhibit GeV invariant mass peak. 

\item {\it $ZZ$ background: }
Since the dominant contribution is from the on-shell $Z$ boson production,
the invariant mass of two charged leptons is close to $m_{Z^0}$. 
The distribution of $p_T^{e^+ e^-}$, i.e. $p_T^{Z^0}$, peaks around 30~GeV. 

\item {\it $Z\gamma^*$ background: }
It provides very similar collider signature as the signal. Originated from
a off-shell photon decay, two charged leptons move in parallel and have a small
invariant mass. The transverse mass of the lepton system and $\met$ does not
peak around $m_{Z^0}$. Unlike the signal, the off-shell photon decays promptly 
and do not travel a long distance inside the detector. 
\end{itemize}
The SM backgrounds, mainly from $WW$ production, overwhelm the signal. 
Making use of the kinematics difference mentioned above, one can impose optimal 
cuts to suppress the SM background. For illustration, we impose the following 
two simple cuts:  
\begin{equation}
m_{e^+ e^-} < 10~\rm{GeV}\quad \rm{and}\quad \cos\theta_{e^+ e^-} > 0.95,
\label{eq:cut}      
\end{equation}
where $m_{e^+ e^-}$ is the invariant mass of lepton-jets and $\theta_{e^+ e^-}$
is the open polar angle between the electron and positron.
Figure~\ref{fig:lhc} shows the $p_T^{e^+ e^-}$ distributions of signal and 
background with the cuts in Eq.~\ref{eq:cut}, which clearly shows that
it is very promising to observe the dark gauge boson signal at the LHC.  
However, one should bear in mind that the above two cuts serve for the purpose
of suppressing the SM background processes. It is crucial to understand 
how well one can measure the energy and momentum of the lepton jets, which is
beyond the scope of this paper and will be presented in the future work.  

\section{Conclusions}
Discovering dark matter would be an undoubted evidence of new physics beyond
the Standard Model. Recently, many excesses were observed in the direct and 
indirect search of dark matter~\cite{Strong:2005zx,
Chang:2008zzr,Adriani:2008zr, Adriani:2008zq,Abdo:2009zk,Bernabei:2008yi}.
In this work we proposed a non-Abelian dark matter model to explain all those 
observed excesses. The model consists of both SM sector and dark sector. 
The latter, gauged under a new $SU(2)$ dark gauge symmetry, contains $SU(2)$
dark gauge boson fields, two triplets dark Higgs fields and two generation 
dark fermion doublets (dark matter candidates). Rather than considering only one
dark matter candidate, we introduced two coexisting stable dark matters:
one is around 500~GeV and the other is around 1300~GeV. Our study shows that
two dark matter scenario naturally fits both PAMELA and Fermi LAT data 
simultaneously. Dark matter annihilation occurs in the dark sector and the 
remanent dark particles (dark gauge boson and dark scalars) decay into 
the SM particles eventually through the kinetic and mass mixing operators 
between the SM and dark sectors. The triplet dark scalars are introduced
to break the dark gauge symmetry and also generate small mass splits between
the two component fields of dark fermion doublets. Such small mass splits
is the key to realize the iDM and XDM scenarios, which explain the INTEGRAL
and DAMA results.  Finally, we explore the interesting collider signature
of the dark gauge boson and dark scalar production at the LHC.
Our simulation analysis indicates that the signals of this model may be detectable at the LHC.

In our model, since the Lagrangian of the dark sector has scalar mass terms, 
there is a dark sector fine tuning problem. In general, such a fine tuning 
problem can be solved by introducing supersymmetry in the dark sector. 
The supersymmetry extension of this model will be especially interesting and 
will be presented elsewhere. 

\begin{acknowledgements}
This work is supported in part by the National Natural Science Foundation of 
China, under Grants No.10721063, No.10975004 and No.10635030. 
Q.~H.~C. is supported in part by the Argonne National Laboratory and 
University of Chicago Joint Theory Institute (JTI) Grant 03921-07-137, 
and by the U.S.~Department of Energy under Grants No.~DE-AC02-06CH11357 
and No.~DE-FG02-90ER40560.
Zhao Li is supported in part by the U.S. National Science Foundation under
Grant No. PHY-0855561.
\end{acknowledgements}
 
\appendix

\section{ANOMALOUS DIMENSION OF THE OPERATOR}\label{sec:operator}
In this appendix, we show the evolution of the dimension-five operator in Eq.~\ref{eq:mixoperator}, since the operator evolution 
can change the theoretical result significantly~\cite{Braaten:1990gq}. 
%To determine the evolution of Eq.~\ref{eq:mixoperator}, we need the anomalous dimension of this operator which 
%can be obtained from the calculation of Feynman diagrams in Fig. 8.
The evolution of the Wilson coefficient $C(\mu)$ of the effective Hamiltonian $\frac{1}{2\Lambda}C(\mu)\mathcal{O}(\mu)$ 
($\mathcal{O}\equiv B_{\mu\nu}\zd^{a,\mu\nu}\hd_2^a$) is determined by the evolution equation
\begin{equation}
\mu\frac{d}{d\mu}C=\gamma_{\mathcal{O}}C,
\end{equation}
where $\gamma_{\mathcal{O}}$ is the anomalous dimension. Calculating the Feynman diagrams in Fig.~6, we can obtain 
\begin{eqnarray}
\gamma_{\mathcal{O}}=-\frac{g_D^2}{24\pi^2}\left(1-n_f \right),
\end{eqnarray}
where $n_f$ is the number of fermion doublet as defined in Sec. III. A.
The solution of the evolution equation (A1) is 
\begin{equation}
C\left(\mu\right)=\exp\left[\int_{g_D(\Lambda)}^{g_D(\mu)}\frac{\gamma_{\mathcal{O}}(g_D)}{\beta(g_D)}dg_D\right]C\left(\Lambda\right).
\end{equation}
Thus, the Wilson coefficient at $M_\zd$ scale becomes
\begin{equation}
C(M_{\mathcal{Z}})=\left(\frac{\alpha_D(M_{\mathcal{Z}})}{\alpha_D(m_1)}\right)^{1/20}\left(\frac{\alpha_D(m_2)}{\alpha_D(\Lambda)}\right)^{-1/16}C(\Lambda).
\end{equation}

\begin{figure}[]
\includegraphics[width=0.8\textwidth]{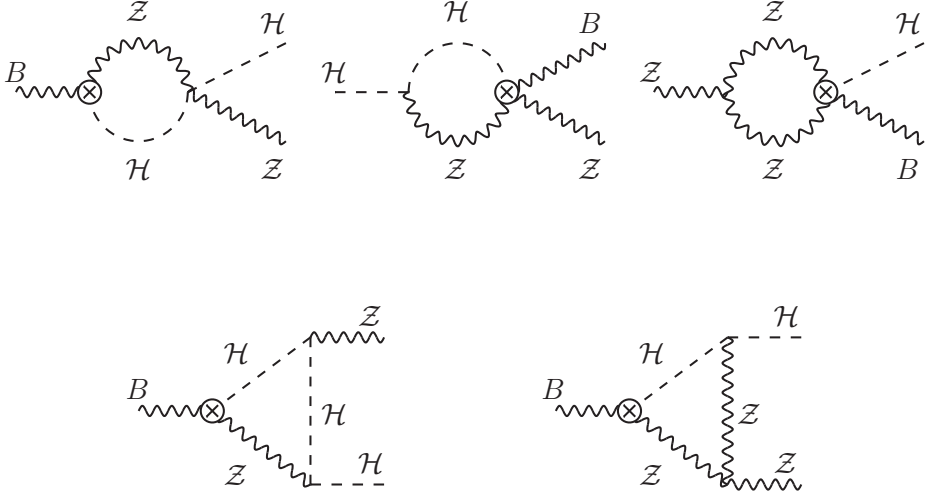}
\caption{The feynman diagrams which contribute the operator anomalous 
dimension. } 
\label{fig9}
\end{figure}

\section{DECAY OF DARK GAUGE BOSONS AND DARK SCALARS}\label{sec:decay}
In this appendix, we discuss the decay properties of dark gauge bosons and dark Higgs bosons.

\subsection{decay of $\zd_1$}
The main decay channels of $\zd_1$ are $\zd_1\rightarrow h_1 h_2$ and $\zd_1\rightarrow h_1 \tilde{h}_3$. 
The total decay width is given by
\begin{equation}
\Gamma_{\zd_1}=\frac{\alpha_DM_{\mathcal{Z}}}{6\sqrt2}\left(1-\frac{2M_{\mathcal{H}}^2}{M_{\mathcal{Z}}^2}\right)\sqrt{1-\frac{2M_{\mathcal{H}}^2}{M_{\mathcal{Z}}^2}}.
\end{equation}
For $M_{\mathcal{Z}}=0.1$GeV, $M_{\mathcal{H}}=0.02$GeV and $\alpha_D(M_{\mathcal{Z}})=0.57$, we have $\Gamma_{\zd_1}\sim0.006$GeV.

\subsection{decay of $\zd_2$}
The main decay channels of $\zd_2$ are $\zd_2\rightarrow \gamma h_2$ and $\zd_2\rightarrow f\bar{f}$ 
($f=e^-,~\nu_{e},~\nu_{\mu},~\nu_{\tau}$). For $\gamma h_2$ channel, the decay width is
\begin{equation}
\Gamma_{\zd_2\rightarrow\gamma h_2}=\frac{\alpha_D\epsilon^2c_W^2M_{\mathcal{Z}}}{12}\left(1-\frac{M_{\mathcal{H}}^2}{M_{\mathcal{Z}}^2}\right)^3.
\end{equation}
The decay width of channel $\zd_2\rightarrow e^+e^-$ is
\begin{equation}
\Gamma_{\zd_2\rightarrow e^+e^-}=\frac{\alpha\epsilon^2c_W^2M_{\mathcal{Z}}}{3}\left(1+\frac{5m_e^2}{2M_{\mathcal{Z}}^2}-\frac{m_e^4}{2M_{\mathcal{Z}}^4}\right)\sqrt{1-\frac{4m_e^2}{M_{\mathcal{Z}}^2}},
\end{equation}
where $\alpha$ is the fine structure constant. The decay widths $\Gamma_{\zd_2\rightarrow\nu_i\bar{\nu_i}},~(i=e,\mu,\tau)$ 
are negligible small. So the total decay width of $\zd_2$ can be written as
\begin{equation}
\Gamma_{\zd_2}=\frac{\alpha\epsilon^2c_W^2M_{\mathcal{Z}}}{3}\left(1+\frac{5m_e^2}{2M_{\mathcal{Z}}^2}-\frac{m_e^4}{2M_{\mathcal{Z}}^4}\right)\sqrt{1-\frac{4m_e^2}{M_{\mathcal{Z}}^2}}+\frac{\alpha_D\epsilon^2c_W^2M_{\mathcal{Z}}}{12}\left(1-\frac{M_{\mathcal{H}}^2}{M_{\mathcal{Z}}^2}\right)^3.
\end{equation}
For $M_{\mathcal{Z}}=0.1$GeV, $M_{\mathcal{H}}=0.02$GeV, $\epsilon=10^{-5}$ and $\alpha_D(M_{\mathcal{Z}})=0.57$, we have $\Gamma_{\zd_2}\sim3\times10^{-13}$GeV.%, so it is detectable at the LHC. Its signal at the LHC is $e^+e^-$ or $\gamma+\displaystyle{\not}E_T$.

\subsection{decay of $\zd_3$}
The main decay channels of $\zd_3$ is $\zd_3\rightarrow\gamma \tilde{h}_3$. The total decay width is
\begin{equation}
\Gamma_{\zd_3}=\frac{\alpha_D\epsilon^2c_W^2M_{\mathcal{Z}}}{24}\left(1-\frac{M_{\mathcal{H}}^2}{M_{\mathcal{Z}}^2}\right)^3.
\end{equation}
For $M_{\mathcal{Z}}=0.1$GeV, $M_{\mathcal{H}}=0.02$GeV, and $\alpha_D(M_{\mathcal{Z}})=0.57$, we have $\Gamma_{\zd_3}\sim1.6\times10^{-13}$GeV.

\subsection{decay of $h_1$}
The main decay channels of $h_1$ are $h_1\rightarrow f\bar{f}f'\bar{f}'$ 
where $f,f'=e^-,~\nu_e,~\nu_{\mu},~\nu_{\tau}$. For $M_{\mathcal{Z}}=0.1$GeV, $M_{\mathcal{H}}=0.02$GeV and 
$\alpha_D(M_{\mathcal{Z}})=0.57$, the numerical result of the decay width shows that it is of order $O(10^{-35}\rm{GeV})$. %It will go out of the galaxy before decaying, so it does not contribute cosmic ray $e^+e^-$ in our calculation. The signal of $h_1$ is missing energy at the LHC. 

\subsection{decay of $h_2$}
The dominant decay channels of $h_2$ are $h_2\rightarrow\gamma f\bar{f}$ ($f=e^-,~\nu_{e},~\nu_{\mu},~\nu_{\tau}$). 
The numerical calculation shows that $\Gamma_{h_2\rightarrow\gamma\nu_i\bar{\nu_i}},~(i=e,\mu,\tau)$ are negligible small. Neglecting the electron
mass and the contribution from diagrams with the $Z^0$ boson in the internal line, the total decay width can be written as
\begin{equation}
\Gamma_{h_2}=\frac{\alpha\alpha_{D}c_{W}^{4}\epsilon^{4}M_\hd}{72\pi}\left[-17\frac{M_\hd^2}{M_\zd^2}+42-24\frac{M_\zd^2}{M_\hd^2}+6\left(1-\frac{4M_\zd^2}{M_\hd^2}\right)\left(1-\frac{M_\zd^2}{M_\hd^2}\right)^{2}{\ln}\left(1-\frac{M_\hd^2}{M_\zd^2}\right)\right].
\end{equation} 
For $M_{\mathcal{Z}}=0.1$GeV, $M_{\mathcal{Z}}=0.02$GeV, $\epsilon=10^{-5}$ and $\alpha_D(M_{\mathcal{Z}})=0.57$, 
$\Gamma_{h_2}\sim4\times10^{-32}$GeV.% It will go out of the galaxy before decaying, so does not contribute to cosmic ray $e^+e^-$ in our calculation. The signal of $h_2$ is missing energy at the LHC. 

\subsection{decay of $\tilde{h}_3$}
$\tilde{h}_3$ decays into $e^+e^-$. The decay width is
\begin{equation}
\Gamma_{\tilde{h}_3}=\frac{\epsilon_{\hd}^{2}M_\hd}{32\pi^2\alpha_{D}}\frac{m_{e}^{2}M_\zd^2}{m_H^4}\left(1-\frac{4m_e^2}{M_\hd^2}\right)\sqrt{1-\frac{2m_e^2}{M_\hd^2}}.
\end{equation}
For $m_H=115$GeV, $M_{\mathcal{Z}}=0.1$GeV, $M_{\mathcal{H}}=0.02$GeV, $\epsilon_{\hd}=10^{-7}$ and $\alpha_D(M_{\mathcal{Z}})=0.57$, the 
decay width of $\tilde{h}_3$ is $1.7\times10^{-35}$GeV. %It will go out of the galaxy before decaying, so does not contribute to cosmic ray $e^+e^-$ in our calculation. The signal of $\tilde{h}_3$ is missing energy at the LHC. 

\bibliographystyle{apsrev}
\bibliography{dark_matter}

\end{document}